\documentclass[aps,physrev,twocolumn,groupedaddress]{revtex4-2}
\usepackage{graphicx}
\usepackage{longtable}
\usepackage{booktabs}
\usepackage[normalem]{ulem}
\useunder{\uline}{\ul}{}

\begin{document}

\title{\textbf{Data-Driven by Design: Building a Reflective Physics Graduate Program}}

\author{Kevin Coldren}
 \email{Contact author: ktc9926@rit.edu}
\author{Diana Sachmpazidi}%
 \email{Contact author: dxssps@rit.edu}
\affiliation{%
  School of Physics and Astronomy, Rochester Institute of Technology, Rochester, NY, 14623 \\
}

\date{\today}

\begin{abstract}

Well-documented research on physics graduate education has demonstrated long-standing issues that hinder equitable student access and participation. Addressing these challenges can be particularly difficult because they are often rooted in entrenched disciplinary and departmental cultures that tend to be rigid and resistant to change. In this work, we aim to cultivate a data-driven culture of cyclic self-reflection and action to proactively identify and address issues that affect student well-being and success in both a new physics and a long-standing astrophysics graduate program within a single institution. Drawing primarily on qualitative data (open-ended survey responses and focus-group interview data) from 15 students in both programs, we collaborated with the program leadership to identify actionable steps for improvement. In this paper, we present findings on student experiences across the two programs and discuss implications for research and practice. More broadly, this work provides a framework for graduate programs seeking to build a data-driven culture that improves student experiences.

\end{abstract}

\maketitle

\section{Introduction}

Graduate programs in Science, Technology, Engineering, and Mathematics (STEM) have a long and well-documented history of high student attrition rates~\cite{Sowell2015} and disparities in access and participation across the categories of race, ethnicity, and gender identity\cite{oakes_chapter_1990,Sowell2015, NSF2018}. Many of these STEM departments feature graduate admissions committees that have faculty with fixed mindsets around the abilities of students, which can negatively impact the participation of women and minoritized racial or ethnic groups in STEM~\cite{dweck_mindset_2016,scherr_fixed_2017}. Graduate STEM programs also often perpetuate a white-male-dominant culture built around competition and individualism~\cite{ong_counterspaces_2018}, causing many students with underrepresented identities in STEM to be socially isolated and academically unsupported~\cite{golde2005role,ali2006dealing}, which in turn can lead to increased attrition~\cite{lovitts2001leaving,ali2007dealing}.

These concerns are particularly pronounced in physics, which remains the  least diverse of all STEM fields~\cite{maries2024towards}. Despite incremental gains, less than 20\% of physics Ph.D. degrees from U.S. institutions were granted to women between 1980 and 2019~\cite{mulvey19}, with the proportion only rising to 21.2\% in 2024~\cite{ncses_sed2024}. This disparity is even more pronounced for African American, Native American, and Hispanic American students. People from these marginalized communities make up approximately 33\% of the U.S. population~\cite{census}, yet only received 3\% of physics Ph.D. degrees in 2019~\cite{mulvey19}. Overall collective attrition rates across physics Ph.D. programs in the United States remain as high as 59\%~\cite{sowell2008phd}. 

In response, several initiatives have sought to increase access and participation of students of these underrepresented identities in graduate physics programs~\cite{apsbridge,igenetwork}. For example, the American Physical Society Bridge Program (APSBP) supports students from diverse backgrounds to bridge the gap between their undergraduate and graduate experiences and reports a retention rate of 92\% among participants~\cite{hodapp2017bridge}. However, research indicates that students in such programs may still encounter resentment and microaggressions within their departments ~\cite{gamez2022graduate}. These findings suggest that while structural interventions can improve access and retention, they do not necessarily transform the underlying departmental cultures that shape students' day-to-day experiences ~\cite{goldberg2024inclusive,cochran_understanding_2025}.

Culture, which refers to a set of shared practices, norms, beliefs, behaviors, and assumptions within an organization \cite{Schein04}, plays a central role in shaping graduate student experiences. Efforts to meaningfully alter departmental culture require deliberate, sustained, and collaborative reflection, and supported by systematic data collection \cite{eckel_taking_2011,kezar_how_2018,sachmpazidi2022changing,sachmpazidi2026psychometric}. Such work constitutes  \textit{second order change}, which addresses multiple dimensions of an organization and unfolds over the long term, as opposed to short, small incremental change in isolated areas (known as \textit{first order change}) \cite{kezar_how_2018,ngai2020}. Despite growing recognition of the importance of culture, many change efforts in physics graduate education continue to solely emphasize structural reforms without fully addressing the cultural dynamics that sustain inequities ~\cite{goldberg2024inclusive}.

A notable example of major change efforts in the field of graduate physics education is a graduate program, where a tragedy within the physics community prompted programmatic reforms~\cite{barthelemy_graduate_2023}. The department paused graduate admissions for a full academic year to reflect and address serious and ingrained issues with the policies and culture of the program~\cite{lenz_graduate_2025}. While this case demonstrates that meaningful change is possible, it also illustrates the reactive nature of many reform efforts, which often occur after significant harm has occurred. Prior research further shows that although physics faculty often express interest in using student data to guide departmental change, responsibility for collecting and interpreting such data is frequently siloed at the individual faculty level ~\cite{sachmpazidi2024recognizing}. As a result, discussions of student data tend to remain informal and rarely lead to coordinated, collective action ~\cite{sachmpazidi2022changing,sachmpazidi2024recognizing}.

Astronomy and astrophysics constitute major subfields within the broader field of physics, though they exhibit some important differences. For example, astronomy and astrophysics tend to have greater gender diversity than physics for both students and faculty in the field~\cite{mulvey19,porter_ivie2019_womeninphysics,barthelemy_educational_2020}. The presence of STEM faculty who identify as women has been shown to increase interest and participation among female students by providing visible role models~\cite{dasgupta_girls_2014}, a dynamic that may also apply in  astronomy and astrophysics. However, minoritized groups, including Hispanic or African-American students remain underrepresented in astronomy or astrophysics graduate programs~\cite{porter_ivie2019_womeninphysics}. Structurally, astronomy or astrophysics graduate programs may be housed within physics departments or as separate departments on the same campus, with the latter arrangement being more common ~\cite{mulvey19}. As a result, physics, astronomy, and astrophysics  are often treated as a single combined domain in the literature (e.g.~\cite{abdurrahman_objectivity_2021,goldberg2024inclusive,sachmpazidi2024recognizing}), despite the distinctions among the fields in many contexts ~\cite{porter_ivie2019_womeninphysics,barthelemy_educational_2020}.

This study examines an effort to cultivate a culture of continuous program improvement within one physics and astronomy department through systematic data collection, collective reflection, and responsive action. The department recently welcomed the inaugural cohort of students into a newly designed physics Ph.D. program, housed alongside an established M.S. program in physics and longstanding M.S. and Ph.D programs in astrophysics. New doctoral programs do not emerge in vaccum; rather, they take shape within preexisting departmental norms, structures, and histories. This context therefore provided a valuable opportunity for researchers and departmental leadership to proactively shape the emerging physics Ph.D. program while it was still in its formative stages.  

At the same time, the presence of the established astrophysics graduate program enables parallel reflection on existing student experiences within the department. In this study, we examined how the structures and culture of the astrophysics program shaped student experiences, both to identify areas for improvement and to surface established strengths that could inform the intentional development of the new physics Ph.D. program. In this way, the astrophysics program served simultaneously as a site for reflection and as a source of cultural and structural insights for proactive program design. 

To support these goals, the researchers designed a study to systematically document graduate student experiences across the department and to share these findings with the departmental leadership in order to collaboratively develop actionable plans for improvement. The work presented in this paper was guided by the following research questions:

RQ1: \textit{How do graduate students describe the supports, challenges, and cultural elements of their doctoral programs, and what areas of strength and improvement do they identify?}

\indent\indent RQ1a: \textit{How do students in the first cohort of the newly developed physics doctoral program describe these experiences?}
  
\indent\indent RQ1b: \textit{How do students in the longstanding astrophysics doctoral program describe these experiences?}
  
\indent\indent RQ1c: \textit{How do student experiences compare between the newly developed physics doctoral program and the longstanding astrophysics program within the same department?}

RQ2: \textit{ How does program leadership respond to findings and collaborate with researchers to create actionable plans that support meaningful change toward improving graduate student experiences and outcomes?}

\section{Literature Review}

\subsection{Graduate Program Design}

Although STEM Ph.D. programs vary across institutions, several structural features are widely characteristic of the graduate student experience. The first year of a STEM Ph.D. program typically involves a combination of graduate coursework, teaching assistant responsibilities, and the initial stages of conducting research under the supervision of a faculty advisor. In physics programs, required coursework commonly includes classical mechanics, classical electrodynamics, statistical mechanics, and quantum mechanics, as outlined in the 2006 Report on Graduate Education in Physics by the Joint AAPT–APS Task Force~\cite{CampbellEtAl2006}.

A major early milestone in most STEM Ph.D. programs is the qualifying exam, which students must pass to advance to Ph.D. candidacy~\cite{mclaughlin_standards_2024}. These qualifying or comprehensive exams take different forms across programs but typically consist of written or oral assessments based on course content, research content, or a combination of both, and are evaluated by a committee of graduate faculty~\cite{mclaughlin_standards_2024}. Because of their high-stakes nature, qualifying exams can function as gatekeeping mechanisms and have been discussed as potentially perpetuating systemic barriers for students with minoritized identities in STEM~\cite{basir_departmental_2024}.

Faculty perspectives on the role of qualifying exams are mixed. In a study by Basir and Burkholder ~\cite{basir_investigating_2024}, physics faculty emphasized the importance of assessing student readiness to succeed in the field of physics while also acknowledging that current exam structures may require substantial revision to accurately make such assessments. More broadly, although graduate program design in physics remains strongly shaped by longstanding traditions, many departments are now pursuing reform efforts aimed at improving student experiences and retention.

\subsection{Financial Support}
Graduate students often face substantial financial pressures related to tuition, housing, and everyday living expenses. In STEM doctoral programs, these costs are frequently mitigated through institutional funding packages that offset tuition and provide a stipend to help cover local living costs~\cite{feizi_satisfaction_2023}. According to the 2023 Survey of Graduate Students and Postdoctorates in Science and Engineering, the three most common primary funding sources for STEM doctoral students were research assistantships (41.2\%), teaching assistantships (23.0\%), and fellowships (15.5\%), while a smaller proportion of students were self-funded (9.3\%)~\cite{nsf_gss2023}.

However, this funding structure differs substantially for master’s students. These sources of support are often not available to STEM graduate students in M.S. programs; since 2017, approximately 70\% of students in STEM M.S. programs have paid for their tuition and living expenses through self-funding~\cite{nsb2023_academicRD,nsf_gss2023}.

Even when funding is provided, financial security is not guaranteed. Zhou and Okahana ~\cite{zhou_role_2019} found that increased departmental financial support is positively associated with graduate student retention, with research assistantship funding showing stronger associations with degree completion than teaching assistantship funding. Nevertheless, students who receive tuition coverage and stipends do not always experience financial stability. For example, a survey of clinical psychology doctoral students found that despite receiving funding toward their degree, over 80\% of respondents reported incurring additional debt to meet living costs~\cite{szkody_financial_2023}. Financial uncertainty may be particularly acute for international students; Laufer \textit{et al.} ~\cite{laufer_invisible_2019} found that some international students reported little to no certainty about their funding status, contributing to adverse mental health outcomes and, in some cases, program attrition. Together, these findings suggest that both the availability and the stability of funding play important roles in graduate student persistence and well-being.

\subsection{Mentoring and Research}
Much of the development of STEM graduate students throughout their time in a program occurs outside of the traditional classroom in the form of mentoring relationships. According to a review of the literature, there is no agreed-upon formal definition of student mentoring~\cite{crisp2009mentoring}. The most common mentoring relationship for STEM graduate students is that with the faculty research advisor, though other mentors can include academic advisors, teaching faculty, or senior peer mentors~\cite{wright-harp2008_mentoring}. These mentoring relationships may be informal or formally structured through the department and typically involve some form of graduate student development through guidance, advice, and feedback~\cite{crisp2009mentoring}.

The research advisor is often the most critical relationship for a graduate student throughout their time in a program, and this single faculty member is usually responsible for developing a student's skills in conducting research, scientific writing and publishing, professional networking, job placement after graduation, and general professional development~\cite{barnes2009role}. The centrality of the research advisor can therefore have a disproportionately large influence on a student's overall graduate school experience~\cite{lyons1990mentor,sverdlik2018phd}. A positive relationship with one's research advisor is associated with higher grades and retention rates~\cite{campbell1997faculty}, whereas a negative relationship can lead to adverse mental health outcomes and attrition~\cite{golde2005role}. The Effective Practices for Physics Programs (EP3) Guide recommends mitigating this risk by implementing a multi-mentoring model, in which departments assign multiple formal mentors to each graduate student, such as a research advisor, academic advisor, and senior peer mentor~\cite{ep3_advising_mentoring_2025}.

Given this central role, the process of finding a faculty research advisor can itself be stressful for graduate students, particularly as they balance coursework and teaching duties in their first year~\cite{verostek2023inequities}. Many factors must be considered when choosing an advisor, including the potential working relationship with the advisor, alignment with the research topics of the group, and the availability of funding~\cite{verostek2024modeling}. If a graduate student determines that the relationship with their chosen advisor is not working as expected, changing to a new advisor and research group can involve several barriers, including funding constraints, personal relationships with mentors and peers, and lost time toward graduation~\cite{verostek2024physics}. Students with minoritized identities have also reported a disproportionate lack of advice and mentoring when selecting a research advisor and group~\cite{verostek2024physics}.

\subsection{Community and Sense of Belonging}
Within graduate education, the social aspect of the graduate school experience is an important indicator of a student's mental health~\cite{gin_phdepression_2021}, as well as their likelihood to persist in their program~\cite{sverdlik2018phd}. First-year graduate students are often at a new institution and in a new geographic location compared to their undergraduate schooling, which can make it difficult to build new relationships and form meaningful social bonds with peers~\cite{apsbridge,hodapp2017bridge}. An overall sense of social isolation in graduate school has been found to be a predictor of attrition from a program in multiple studies~\cite{ali2006dealing,ali2007dealing,gardner_contrasting_2010,lovitts2001leaving}.

In particular, these initial social barriers can be even more pronounced for international students in U.S. graduate programs ~\cite{zhang_predictors_2011,rodriguez_social_2024}. Beyond international status, social isolation also tends to be more pronounced among students with underrepresented identities in the field of physics, such as women and people of color~\cite{johnson2017common,scherr_isolation_2020}, as well as members of the LGBTQ+ community~\cite{barthelemy2022lgbt+}. Bridge programs are designed to help these underrepresented groups succeed in STEM programs; however, these students may still face social barriers such as microaggressions or resentment from peers~\cite{gamez2022graduate}.

Given these challenges, a sense of belonging has a large effect on student persistence in their schooling and careers in STEM, with students from minoritized identities experiencing more negative effects in this area according to Douglas \textit{et al.}~\cite{douglas2024importance}. Johnson \textit{et al.} found that departments have the ability to help develop a sense of belonging among their graduate student body by taking actions such as employing research-based teaching methods and actively addressing and rejecting microaggressions toward students with minoritized identities~\cite{johnson2017common}. Departments that support the development of a sense of community, for example, by providing common workspaces for students, have been shown to have higher graduate program completion rates~\cite{zhou_role_2019}. Together, this body of work highlights the central role that social integration and departmental climate play in shaping graduate student persistence and well-being.

\subsection{Professional Development}
Professional development for graduate students involves both exploring potential career paths and building the skills necessary to enter those fields. Multiple surveys of physics graduate students have identified professional development as one of the most lacking forms of departmental support in physics programs ~\cite{sachmpazidi2021departmental,sachmpazidi2025beneath}. In many cases, doctoral advisors serve as the primary (or only) source of professional development for students within their research groups~\cite{barnes2009role}.
However, advisor-centered models may leave important gaps. A 2021 study by Ganapti and Ritchie ~\cite{ganapati_professional_2021} in STEM Ph.D. programs found that while students commonly developed research, writing, and teaching skills, significant deficiencies remained in the professional skills and networking needed to pursue careers outside academia. Even for students intending to remain in academia, additional preparation may be necessary. For example, Stowell \textit{et al.} ~\cite{love2015transforming} argue that increased training in evidence-based pedagogy is essential for preparing the future education workforce, particularly for Graduate Teaching Assistants (GTAs).

More broadly, effective departmental professional development has been linked to positive student outcomes. O'Meara \textit{et al.} ~\cite{omeara_by_2014} found that robust professional development structures can enhance students’ sense of agency over their career trajectories and may also contribute to improved program retention. Together, these findings suggest that comprehensive department-level professional development is a critical component of graduate student success.

\subsection{Graduate Student Mental Health}

Graduate student life in STEM can be highly stressful and detrimental to students’ mental well-being, and departmental culture has been shown to be a significant contributor to these negative outcomes~\cite{bork_exploring_2022}. A typical first-year STEM Ph.D. experience requires balancing coursework, teaching duties, and research responsibilities, which can make maintaining a healthy work–life balance difficult~\cite{martinez_striving_2013}. Consistent with these structural pressures, a study by Evans \textit{et al.} found that over 80\% of STEM Master's and Ph.D. students surveyed self-reported moderate to severe stress levels, and over 70\% reported moderate to severe burnout~\cite{evans_stress_2018}.

Field-specific evidence further highlights the severity of this issue. In physics and astronomy, Rydberg found that graduate students reported anxiety and depression at rates five to seven times higher than the general population~\cite{banner2025rydberg}. Academic burnout - characterized by Allen \textit{et al.}  ~\cite{allen_graduate_2022} as involving “exhaustion, cynicism, and inefficacy” - has been associated with a higher likelihood of attrition among biomedical doctoral students~\cite{nagy_burnout_2019}. Importantly, these mental health challenges are not experienced equally across student populations. Underrepresented groups in STEM graduate programs may face more acute struggles; for example, neurodivergent students may experience elevated anxiety and burnout due to pressure to mask traits that diverge from departmental norms~\cite{syharat_experiences_2023}.

\section{Research Methods}

\subsection{Educational Context}
\begin{figure*}[t]
\includegraphics[width=0.85\linewidth]{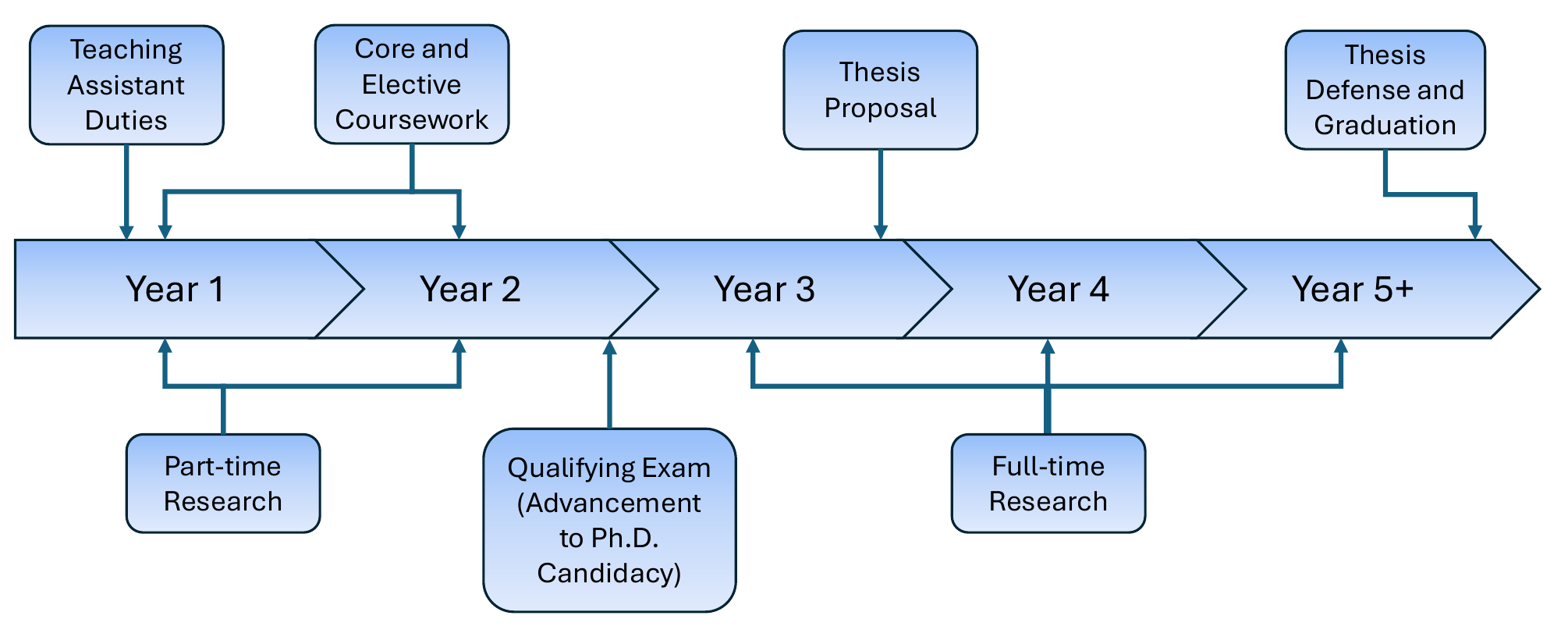}
\caption{\label{fig:PhD_Timeline} The timeline for a typical student in the new Physics Ph.D. program discussed in this paper.}
\end{figure*}

This study was conducted at a private R2 university in the United States. The physics department offers longstanding M.S. and Ph.D. programs in Astrophysics, as well as an M.S. program in Physics; the Physics Ph.D. program examined in this study is the department’s newest graduate offering. The program is designed as a five-year pathway, 
and a typical student progression is summarized in Fig.~\ref{fig:PhD_Timeline}.

During the first year, students balance graduate coursework, research under the supervision of a faculty advisor, and teaching assistant responsibilities. Required core coursework includes standard graduate-level subjects (e.g., Quantum Theory and Electrodynamics). Students who earn a final grade of B or better in these courses may waive the written comprehensive exam associated with the relevant subject area. In addition to the core, students complete four elective graduate courses aligned with their physics subfield and a two-semester graduate seminar sequence focused on faculty research, program milestones, and research skill development.

Students indicate a preferred research advisor during the application process, and faculty availability and funding are considered in admission decisions. Under advisor supervision, students begin research in their first year. The program includes three primary research milestones: the qualifying exam, the thesis proposal, and the dissertation defense. The qualifying exam, taken at the end of the second year, requires a written report based on a master’s-level research project and an oral defense before a faculty committee; successful completion advances students to Ph.D. candidacy. Candidates later present a dissertation proposal and timeline to their committee, which reconvenes to evaluate the completed dissertation and final public defense.

The first cohort of the Physics Ph.D. program consisted of six students, most of whom entered with M.S. degrees from other institutions. Because many transferred with graduate course credits, the graduate seminar sequence served as the primary shared coursework experience. All students began the program with identified research advisors, and most were supported by teaching assistantships in the first year, with some transitioning to research funding by the second semester.

\subsection{Data Collection}

The research team conceptualized this study and contacted departmental leadership to determine whether the program was interested in participating and willing to consider potential action items informed by the findings. After obtaining approval for human subjects research from the university Institutional Review Board (IRB), the department chair and program directors provided the researchers with the email addresses of all graduate students in the physics department. 

Two types of data were collected: qualitative focus group data (the primary source of evidence) and quantitative survey data using the Aspects of Student Experiences Scale (ASES)~\cite{sachmpazidi2021departmental} which was administered through Qualtrics. The survey was used to hear from as many students as possible, recruit a subset of participants for follow-up focus groups, and to provide baseline context situating this program relative to national ASES samples. The ASES survey tool includes 31 Likert-scale items (five-point “Strongly Disagree” to “Strongly Agree”) measuring graduate student perceptions across four factors: Social and Academic Integration, Mentoring and Research Experiences, Professional Development, and Financial Support \cite{sachmpazidi2021departmental}. 

Qualitative data were collected through open-ended questions on the ASES survey and focus groups with first-year graduate students. The open-ended survey questions elicited students' perspectives on program strengths, challenges, and overall experiences. The primary qualitative dataset, however, consisted of focus group discussions with first-year graduate students. All focus groups followed a common 90-minute semi-structured protocol that began with broad questions about students' experiences in the program and then explored topics identified in prior graduate education research. These topics included social integration, mental health and well-being, coursework, research and mentoring experiences, professional development, financial support and off-campus living, and teaching assistantship experiences. 

\subsection{Participants}
The number of student respondents to the ASES from each of the four graduate programs are shown in Table~\ref{table:Grad Programs}, along with the total number of graduate students enrolled in the programs at the time of data collection. The distribution of the years that each respondent had been in their program at the time of data collection is shown in Table~\ref{table:Year Distribution}. Both the quantitative and qualitative data from these survey responses were used for the analysis done in this paper. 

Two focus groups were conducted in-person and were video and audio recorded. Participation was limited to first-year students to ensure better comparison with the inaugural cohort of the newly established Physics Ph.D. program. The first of these interviews included four out of six first-year Physics Ph.D. students. The second included two out of ten first-year Astrophysics Ph.D. students and one out of three Astrophysics M.S. student. These groups were combined because the Astrophysics programs share similar first-year structures and experiences. Demographic information is not reported to protect participant anonymity given the small sample size.

\subsection{Data Analysis}

For the quantitative component, responses to the 31 Likert-style ASES items were converted to a numerical scale, with “Strongly Disagree” assigned a value of 1 and “Strongly Agree” assigned a value of 5. Using the validated ASES factor structure, items were grouped into four constructs - Mentoring and Research Experiences (MRE), Professional Development (PD), Social and Academic Integration (SAI), and Financial Support (FS) - and averaged to produce factor-level scores~\cite{sachmpazidi2021departmental}. Mean values and associated standard deviations were calculated for each factor. 

Given the exploratory role of the quantitative data in this study and small sample size, analyses were conducted descriptively. Factor means were used to situate this department’s results relative to prior uses of the ASES in physics programs~\cite{sachmpazidi2021departmental}. To contextualize the magnitude of the differences between this department and previous studies \cite{sachmpazidi2021departmental}, standardized mean differences using Hedges’ $g$ \cite{hedges1982estimation} were calculated and descriptively interpreted. All quantitative analyses were conducted using R.

The primary analysis for this study was qualitative. Focus group recordings were transcribed verbatim and analyzed using a thematic analysis approach~\cite{cooper_thematic_2012} informed by prior graduate education research and the constructs measured by the ASES. Initial coding combined deductive codes aligned with the four ASES factors and inductive codes that emerged from the data. Codes were iteratively refined through constant comparison across transcripts to identify salient themes in first-year graduate student experiences. The written responses to the qualitative questions in the ASES survey were also coded using the final codebook and added to the overall qualitative data set used for thematic analysis. Representative excerpts were selected to illustrate each theme in the Results section.

Themes identified for the Physics and Astrophysics Ph.D. programs were shared with faculty leadership (program director and assistant program director, respectively) in one-hour meetings with the authors. At the request of program leadership, follow-up meetings were held in which the authors presented prospective action items for each program and engaged in discussions regarding the feasibility of implementing the suggested changes.

\subsection{Quantitative ASES Survey Results}
Because the primary aim of this study is to explore participants' experiences through qualitative data, survey results are presented here to provide contextual background for the findings. Given the small program size, these results should be interpreted as descriptive rather than inferential. 

All graduate students in the physics department were invited to complete the ASES survey via Qualtrics. Seventeen responses were complete and included in the analysis (27\% response rate). Respondents represented the Physics Ph.D., Physics M.S., and Astrophysics Ph.D. programs, with no responses from the Astrophysics M.S. program. Program-level participation is summarized in Table~\ref{table:Grad Programs}.

\begin{table}
\centering
\begin{ruledtabular}
\caption{\label{table:Grad Programs}%
ASES respondents by graduate program and total program enrollment at the time of data collection.}
\begin{tabular}{lcc}
Graduate Program & ASES Responses & Total Students \\
\hline
Physics Ph.D.        & 3 & 6 \\
Physics M.S.         & 5 & 7 \\
Astrophysics Ph.D.   & 9 & 44 \\
Astrophysics M.S.    & 0 & 6 \\
\end{tabular}
\end{ruledtabular}
\end{table}

\begin{table}
\centering
\begin{ruledtabular}
\caption{\label{table:Year Distribution}%
ASES respondents by year in their graduate program. N/A values represent program years that did not have any students actively enrolled at the time of data collection, or are beyond the point of intended graduation for a specific program.}
\begin{tabular}{lcccc}
Graduate Program & Year 1 & Year 2 & Year 3 & Year 4+ \\
\hline
Physics Ph.D.        & 3 & N/A & N/A & N/A \\
Physics M.S.         & 3 & 2 & N/A & N/A   \\
Astrophysics Ph.D.   & 3 & 2 & 1 & 3\\
Astrophysics M.S.    & 0 & 0 & N/A & N/A \\
\end{tabular}
\end{ruledtabular}
\end{table}

Survey items were grouped based on previously identified factors ~\cite{sachmpazidi2021departmental}. Table~\ref{tab:ASES_comparison} compares factor means from this department with prior results from a 2021 study where the ASES was given to a total of 397 physics graduate students across 19 U.S. institutions~\cite{sachmpazidi2021departmental}.

\begin{table}
\centering
\begin{ruledtabular}
\caption{\label{tab:ASES_comparison}%
ASES factor comparisons across datasets. Hedges’ $g$ values represent standardized differences between this study (N = 17) and the 2021 study conducted across 19 institutions (N = 397)~\cite{sachmpazidi2021departmental}.}
\begin{tabular}{lccccc}
& \multicolumn{2}{c}{2021 Study}
& \multicolumn{3}{c}{Current Study}  \\
\cline{1-6}
Factor & Mean & SD & Mean & SD & $g$ 
 \\
\hline
MRE & 3.72 & 0.70 & 4.21 & 0.42 & 0.71  \\
PD  & 2.33 & 0.75 & 2.79 & 0.75 & 0.61  \\
SAI & 2.78 & 0.66 & 3.28 & 0.59 & 0.77  \\
FS  & 3.87 & 1.06 & 3.38 & 1.03 & -0.46 \\
\end{tabular}
\end{ruledtabular}
\end{table}

Overall, the pattern of factor averages in this department has some moderate differences when compared with national results. Both Mentoring and Research Experiences (MRE) and Social and Academic Integration (SAI) had moderate-to-large positive effect sizes, suggesting generally positive student perceptions in these areas when compared to their counterparts at other institutions \cite{sachmpazidi2021departmental}. Professional Development (PD) also had a moderate positive effect size, though the mean factor value remained below the neutral midpoint (3), consistent with prior work indicating comparatively weaker support in this domain. Financial Support (FS) had a small-to-moderate negative effect-size, indicating weaker support in this area for the current program. However, this is likely explained by the inclusion of Physics M.S. respondents, whose tuition is not covered by the department at the institution under study. 

The relative ordering of factors was also similar to national patterns, with one notable difference: Financial Support (FS) ranked second rather than first. This shift is consistent with the inclusion of the aforementioned Physics M.S. respondents. Program-level comparisons presented in Appendix Tables~\ref{tab:ASES_comparison_MS_PhD} and \ref{tab:ASES_comparison_Phys_Astro} further illustrate these differences.

These quantitative survey results offer some context for the climate and culture in the department under study. The collective responses indicate an overall typical experience for these graduate students, with some aspects being above average when compared to student views in other programs. While the response rate for this survey was typical, the total number of respondents (N = 17) is not sufficient to do any more in-depth statistical analysis. We are also not able to draw significant conclusions about the department under study based on the collected quantitative data alone. The qualitative analyses that follow provide a more detailed and complete picture of student experiences in this program.

\subsection{Trustworthiness}

To enhance the trustworthiness of the qualitative analysis, several strategies were employed \cite{creswell2023research}. The primary researcher (first author), a graduate student, conducted the data collection and analysis in close consultation with the faculty advisor (second author) whose research expertise is in graduate education and who developed the ASES instrument used in this study. The primary researcher met weekly with the advisor throughout the project to review analytic decisions, coding development, and emerging interpretations.

Peer debriefing was also used to strengthen analytic rigor. The research team shared developing codes, themes, and interpretations with additional science education researchers to solicit critical feedback and alternative perspectives. This process helped refine emerging themes.

Member checking was conducted with study participants to ensure that the reported themes accurately reflected their experiences. Participants were invited to review the draft manuscript and provide feedback on the interpretation of their perspectives. 14 of the 17 students in our data set responded to this request, and the majority of these students found no issues with our representations and interpretations of their experiences. Five of these students asked for minor edits around their quotes, which were mostly to address issues pertaining to potentially identifying information. The requested changes were made to the manuscript in the editing process. Together, these strategies support the credibility and trustworthiness of the findings.

\subsection{Positionality} 
The first author’s positionality as a graduate student provided both opportunities and potential limitations for this work. This status may have helped create a more comfortable environment for focus group participants to share perspectives and also informed the interpretation of graduate student experiences. The second author is a faculty member with expertise in departmental change and graduate education and provided ongoing methodological and interpretive guidance throughout the project.

Both authors recognize that their positions within the academic environment under study shape how the data were interpreted. The research team engaged in ongoing reflective discussion during data analysis and interpretation to critically examine assumptions and support credible qualitative findings.

\section{Qualitative Results}
The themes that emerged from the thematic analysis of the collected qualitative data from the students in both Ph.D. programs fell into two categories: Valuable Program Aspects and Needed Program Changes or Improvements. These themes are further separated by the two programs, and are examined in more detail in the following  four sections. The themes that fit in each of these broader categories are summarized below in Table~\ref{tab:Themes Chart}. 

\begin{table*}[]
\caption{A summary table of the themes that emerged from thematic analysis of the collected qualitative student data. Themes for both programs were sorted into student perspectives on Valuable Program Aspects and Needed Program Changes or Improvements.}
\label{tab:Themes Chart}
\begin{tabular}{@{}cccc@{}}
\toprule\toprule
\begin{tabular}[c]{@{}c@{}}Valuable Program Aspects: \\ Physics Ph.D.\end{tabular} &
  \begin{tabular}[c]{@{}c@{}}Needed Changes or \\ Improvements: \\ Physics Ph.D.\end{tabular} &
  \begin{tabular}[c]{@{}c@{}}Valuable Program Aspects: \\ Astrophysics Ph.D.\end{tabular} &
  \begin{tabular}[c]{@{}c@{}}Needed Changes or \\ Improvements: \\ Astrophysics Ph.D.\end{tabular} \\ \midrule
\begin{tabular}[c]{@{}c@{}}Positive and Supportive\\ Atmosphere Within the \\ Overall Physics Department\end{tabular} &
  \begin{tabular}[c]{@{}c@{}}Weak Sense of Community \\ Between Peers in the \\ First Ph.D. Student Cohort\end{tabular} &
  \begin{tabular}[c]{@{}c@{}} The Program Features a\\  Welcoming Atmosphere and \\ Strong Social Connections \\ Within the Student Body\end{tabular} &
  \begin{tabular}[c]{@{}c@{}}Sense of Disconnection \\ Within the Program \\ After the First Year\end{tabular} \\
\begin{tabular}[c]{@{}c@{}}Physics Ph.D. Program \\ is Flexible and Allows \\ for the Transfer of Credits\end{tabular} &
  \begin{tabular}[c]{@{}c@{}}\textit{}\\Lack of Professional \\ Development Opportunities, \\ Especially Outside \\ of Academia\end{tabular} &
  \begin{tabular}[c]{@{}c@{}}\textit{} \\Department Support Structures \\ Toward Academic and \\ Professional Development \\ for Students\end{tabular} &
  \begin{tabular}[c]{@{}c@{}}Research Experiences are\\ Varied for Students \\ in the Program\end{tabular} \\
\begin{tabular}[c]{@{}c@{}}Wide Variety of Collaborative \\ and Interdisciplinary \\ Research Opportunities\end{tabular} &
  \begin{tabular}[c]{@{}c@{}}\textit{}\\Students Found That \\ Required Coursework \\ Was Generally Not \\ Relevant to Their Research\end{tabular} &
  \begin{tabular}[c]{@{}c@{}}Many Courses Were \\ Described as Well-Designed \\ and Relevant to Research\end{tabular} &
  \begin{tabular}[c]{@{}c@{}}Some Courses Viewed \\ as Not Engaging \\ or Excessively Difficult\end{tabular} \\
\begin{tabular}[c]{@{}c@{}}Supportive Relationships \\ with Mentors and \\ Research Advisors\end{tabular} &
  \begin{tabular}[c]{@{}c@{}}\textit{}\\Students Wanted More \\ Guidance with Finances \\ and Benefits, Especially \\ for International Students\end{tabular} &
  \begin{tabular}[c]{@{}c@{}}\textit{}\\Faculty and Program \\ Leadership are Generally \\ Supportive and Respectful \\ of Student Mental Health\end{tabular} &
  \begin{tabular}[c]{@{}c@{}}\textit{} \\Several Logistical \\ Barriers and Financial \\ Burdens That the \\ Department Could Mitigate\end{tabular} \\
N/A &
  \begin{tabular}[c]{@{}c@{}}\textit{}\\Desire for More Clarity \\ on Program Milestones \\ and Logistics\end{tabular} &
  N/A &
  N/A \\ \bottomrule\bottomrule
\end{tabular}
\end{table*}

\subsection{Valuable Program Aspects: Physics Ph.D.}

\subsubsection{Positive and Supportive Atmosphere Within the Overall Physics Department}
Throughout the focus group interview with the Physics Ph.D. students, there was a general theme of feeling welcomed  and supported by the overall culture within their program and the department as a whole. While the students did not describe any specific actions taken by the department to make them feel a sense of belonging, they did specifically mention their interactions with teaching faculty, their research advisors, and department support staff. When discussing the culture in the department, one student said:

\begin{quote}
     ``I think the culture here is really nice, compared to the old ones [I've been to]. It has a really supportive culture for the most part." - Allen
\end{quote}

Here, Allen describes his overall positive feelings toward the culture within the physics department in particular. When discussing the academic diversity of the campus as a whole, another student said:

\begin{quote}
    ``Compared to my previous school, the campus is much more diverse - lot of different programs, lot of different students from a lot of different backgrounds. I love it." - Dawn
\end{quote}

Allen and Dawn both compare their perspectives on the culture in their program to those of the graduate institutions they have previously attended, which was a common point of discussion in the focus group interview. When similarly expressing his opinions on the culture within the department, Chris said:

\begin{quote}
    ``I think the department is shockingly healthy for a physics department. From a lot of my experiences and from what my friends at [other] grad schools have said, it just seems like there's a really good air of cooperation [..], teamwork, [and] actually trying to help each other." 
    - Chris
\end{quote}

Chris is expressing his strong positive feelings toward the culture of the physics department, and is also comparing the experiences in other programs in a similar manner to Allen and Dawn. Chris also appears to hold the opinion that physics departments are not usually as cooperative and supportive as his current one, further cementing the value he has for this aspect of the program.  

\subsubsection{Physics Ph.D. Program is Flexible and Allows for the Transfer of Credits}
The interviewed students expressed an appreciation for the many flexible structures within the Physics Ph.D. program. The first of these structures was the ability to transfer graduate credits from previous institutions, shown in a qualitative survey response:
 
\begin{quote}
    ``I also appreciate the large transfer of previous graduate research and class hours that shortens the time needed to get my degree significantly.'' - Allen
\end{quote}

Another student echoed with this sentiment during the focus group interview:

\begin{quote}
    ``I like the fact that [the program] is flexible. If you're coming in with your master's, the program head talks with you [and says]: `you have these many courses, so you don't have to take [them] again.' So depending on your background, they customize and tailor the program for you." - Dawn
\end{quote}

These student quotes demonstrate the high value placed on saving time toward graduation.

The discussion on program flexibility also centered on the elective courses that the students could take toward their degree requirements. While completing 4 out of 5 core courses for the Physics Ph.D. does not offer much variety for students, there are significantly more options for students when choosing classes to count toward their 4 required electives:

\begin{quote}
    ``Coursework (how to meet the course requirements, which courses to take, etc.) is flexible and [able to be personalized].'' - Dawn
\end{quote}

These elective courses are often much more aligned with the research interests of an individual student, as shown in Chris's thoughts:

\begin{quote}
    ``I think [flexibility is] a big strength as well. [...] I think that is really important and makes for a more personalized sort of grad school experience.'' - Chris
\end{quote}

The above quotes illustrate an overall perception among the Physics Ph.D. students that many aspects of their program are versatile and can be customized to meet their goals and future plans, and that this is a major strength of the program.  

\subsubsection{Wide Variety of Collaborative and Interdisciplinary Research Opportunities}
All four of the interviewed students had research advisors picked out before starting the program, and thus were involved in research throughout their first year. These research opportunities were praised by the students as interdisciplinary and collaborative. When asked to discuss strengths of the Physics Ph.D. program, the following exchange occurred:

\begin{quote}
``Collaboration for sure, not just within the [physics] department, but within the college of science as a whole. I just feel like it's one big group of people working on various research topics, and sharing expertise with each other.'' - Chris
\end{quote}

Brandon also gave some examples of his collaboration opportunities, with both seeing these opportunities to work with outside lab groups and institutions as unique and advantageous for their research experience. The students in the Physics Ph.D. program had the opportunity to work in laboratory groups that study a wide variety of cross-disciplinary fields. This wide variety of available research topics was also evident in the diverse set of invited guest speakers that presented at department colloquia throughout the academic year:

\begin{quote}
``I like the different speakers who come in [for] the colloquia [...], even those are interdisciplinary [people] we get to hear from, a lot of people from many different fields.'' - Dawn
\end{quote}

Dawn, Chris, and Brandon all placed significant value on learning from and working with people from many different perspectives, and they see their program as having an emphasis on these opportunities for their students.  

\subsubsection{Supportive Relationships with Mentors and Research Advisors}
The research experience was integral to the first year experience for the members of the Physics Ph.D. cohort. The interviewed students all had designated research advisors and were involved in their laboratory groups and projects from the beginning of their first year in the Physics Ph.D. program. Students indicated the faculty they were interested in working with on their applications, and these faculty were in turn a key part in bringing these students into the program:

\begin{quote}
    ``One of the reasons I did my Ph.D. [was] because of my advisor. I talked to him [before the program], and he seemed [to be] very caring about his students.'' - Brandon
\end{quote}

Chris and Dawn also cited his mentor as a major reason for choosing this program for his studies, showing how central research advisors were in building this first program cohort. The mentoring relationships that the students formed with their research advisors in their first year were described as positive and supportive. The students felt that their advisors held them to high standards of work and developing skills, while also being understanding of maintaining a healthy work-life balance with the other aspects of graduate student life. Faculty and mentor support of mental health was specifically mentioned by several of the interviewed students, with Chris saying:

\begin{quote}
    ``I feel like I am more able to talk with my professors here. If I'm having a problem, I feel like I can actually address that with them instead of beating around the bush." - Chris
\end{quote}

Here Chris is favorably comparing faculty attitudes in his current program to those that he has attended in the past, emphasizing their accessibility, openness, and understanding of mental health struggles. 

\subsection{Needed Program Changes or Improvements: Physics Ph.D.}

\subsubsection{Weak Sense of Community Between Peers in the First Physics Ph.D. Student Cohort}
While the interviewed students cited a healthy and positive overall culture in the physics department, they also felt that there was not a strong sense of community between the members of the first Physics Ph.D. cohort. The ability to transfer graduate credits from other institutions meant that most of the first year students did not have common course time or assignments to act as natural community builders, and the department did not hold many social events for the graduate students to act as an offset. This sentiment is shown in the following exchange: 

\begin{quote}
    Dawn: ``Because I don't take classes I get to see everybody only during the grad seminars, so I'm thankful [...] that I get to see everybody, and at least I know who is in the program. Outside of that, I see very few people."

    Chris: ``This is why I'm advocating for more events. Outside of class time or grad seminar, this interview is more direct interaction with my peers than is otherwise fostered within the department, and I would like to change that." 
    
\end{quote}
These quotes show that Dawn and Chris put a lot of importance on the social aspects of being in graduate school, and want to see improvements for themselves and future students. They also show how common it can be for graduate students to feel isolated, even though they are members of research groups upon starting their program in this case.

When comparing their own experiences to those of other graduate students in the physics department, the Physics Ph.D. students perceived the Astrophysics students as having a stronger social network, as shown in the following quote:

\begin{quote}
    ``[The Astrophysics students] all have classes together, and none of us really have classes like that. But I think those [social] events would help with that a lot, so that we actually see each other [in a structured way].'' - Allen
\end{quote}

Allen recognizes a sense of community within the Astrophysics programs that he perceives as aided by departmental structures, and would like to see that for the Physics Ph.D. students as well.

Physical space was also a significant factor that negatively influenced the formation of social bonds between the Physics Ph.D. students. The physics department and its associated faculty laboratory spaces are spread out between at least 3 different academic buildings on campus, meaning there was a lack of common gathering areas for social interactions. Department leadership made a concerted effort to find a dedicated space on campus with desks for graduate students, but were not able to do so before the academic year began. When asked if physical separation had any effect on the social aspect of the program, one student said:  

\begin{quote}
    ``Yeah, I think it does. If people are gathering in [the other building] to work on stuff or just hang out, I'll really not have much access to that. If it was here [in this building], then I'd be more aware of it. [...] The fact that the shared space is on the other side of campus just sort of makes it not worth it." - Chris
\end{quote}

Chris here sheds some light on how much of the social experience for graduate students is entirely optional for each individual in the program, or considered of secondary importance to research or coursework. Barriers like lack of time or physical space were cited by students as deterrents toward interactions with their peers or department faculty outside of what is required of them. 

\subsubsection{Lack of Professional Development Opportunities, Especially Outside of Academia}
When prompted to discuss the professional development opportunities available from the department in their first year, the Physics Ph.D. students felt that this area was lacking overall. Some class time in the graduate seminar courses was devoted to skills like reading and writing academic papers, writing a resume or CV, and giving oral presentations. However, the students overall felt that these topics were mostly focused around careers in academia, rather than exploring careers in industry or other fields. One student summarized these feelings while discussing professional development:

\begin{quote}
    ``People around us [in our program] aren't necessarily people who did their entire career in industry. And if that's something I want to do, I don't feel like I have the same sort of options available to discuss those things, and they don't necessarily know the things that industry is looking for right now. '' - Chris
\end{quote}

The above quote shows that Chris is interested in working in an industry position after graduating, but feels that his program is at least somewhat disconnected from the current state of career fields outside of academia. 

One student also described the idea of having a personalized professional development plan offered through the department as a desired program feature for the future:

\begin{quote}
    ``I would like to see personalized professional development, like a mentor or somebody to talk to who [...] can make you aware of what opportunities are out there, what career options you have, what can you do with your degree." - Dawn
\end{quote}

Dawn has less clarity on her future career when compared to Chris, and recognizes not only the need for guidance, but also the opportunity to explore options with a mentor. Dawn also specifically thinks this mentor should be a dedicated professional separate from the research advisor, who already fills a wide variety of roles in the development of graduate students.

Some department events in the category of professional development were praised, such as a particular colloquium speaker given by an alumni from this institution who had made a career in the pharmaceutical industry after studying physics:

\begin{quote}
    ``I thought the industry speaker this semester was really good, because he was also an alumnus from here. He was someone who [...] got skills from this place, and then went off and was successful in industry. So how do you do that? I would like to meet more people like that and hear how they did it." - Chris
\end{quote}

Here Chris is showing the role that he thinks the department could be playing in bridging the gap between graduate students and industry professionals via alumni from the same programs.  %

\subsubsection{Students Found That Required Coursework Was Generally Not Relevant to Their Research}
While most of the interviewed students avoided coursework via transferring graduate credits, some of them found the required core physics courses they did take to not be very relevant to their research, especially since they cover similar content to their undergraduate physics courses. When discussing core courses, one student said:

\begin{quote}
    ``I think there's too [few] courses that's offered here for graduate requirements. You have to pick four out of five courses, which \textit{none} of them are related to what I'm doing, and all of them are like a slight upgrade from undergrad courses." - Brandon
\end{quote}

Brandon is specifically expressing frustration with core required courses rather than elective courses, showing emotion in emphasizing his perception that they lack relevance to his research while still taking up much of his time. When describing 
required graduate physics courses, another student said:
\begin{quote}
    ``I think a lot of the graduate curriculum is like, `let's do this undergrad physics with much more complicated math'. [For example], Jackson E\&M is just Griffiths, but with terrible math." - Allen
\end{quote}

Allen is describing a perceived overlap in content between upper division undergraduate and graduate physics course content. Allen is also expressing the common sentiment among the students interviewed here that a graduate electrodynamics course based on \textit{Classical Electrodynamics} by John David Jackson \cite{jackson2021classical} can be an exceedingly difficult and time-consuming experience. 

The conversation with the students also had some focus on the required graduate seminar courses, with the students expressing that they did not get a lot of value out of the time spent on them. These sentiments are exemplified here:

\begin{quote}
    ``In my opinion, they seem like they don't really know what they want to do with [the graduate seminar courses], and they seem almost like a waste of time.'' - Allen
\end{quote}

Allen is describing the common perception that the graduate seminar courses lack cohesion in their content. There was also a sentiment that the first seminar in the fall semester was largely designed around helping students to find a research advisor, which was not needed for the members of the first Ph.D. cohort, who all had advisors chosen before starting the program. This course feature appeared more tailored to the Physics M.S. students who also took the graduate seminars.

\subsubsection{Desire for More Clarity on Program Milestones and Logistics}
Several of the students expressed confusion about department policies and logistics. Some of this lack of clarity came from department communications before entering the program, specifically around TA opportunities and similar financial issues:

\begin{quote}
    ``There were some things that I was told before coming in, that were untrue or that the department was less sure of than they presented initially. One was the availability of TA work during the Summer ([seemed] like a definite opportunity before entering and then being unlikely after joining).'' - Allen
\end{quote}

Allen's quote is part of a larger feeling of uncertainty of financial security past their first year that was shared among the interviewed students. 

Other problems with communications concerned program milestones, such as a timeline and requirements for a student's qualifying exam, or the subsequent thesis proposal and defense. The students overall felt that this lack of clarity could be solved with a graduate student handbook, as shown in the following exchange:

\begin{quote}
Allen: ``It's probably because [the program] is new - I think it's the handbook. [...] There's no guidebook on what you need to do to get the Ph.D. - something you can just reference, that would be good to have."

Chris: ``Yeah, right now it just feels totally open ended. I have no idea where I'm going to be a year from now. Like I'm here, obviously, but what will I be doing? I have no idea."
\end{quote}

Here Chris is indicating that he plans to continue in his program, though he has unease surrounding the particulars of funding and role that he will be filling. These student sentiments show that department transparency appears to be lacking in some areas for the students in the Physics Ph.D. program.  

\subsubsection{Students Wanted More Guidance with Finances and Benefits, Especially for International Students}
Students often found themselves confused around finances and benefits received through working for the department in their first year. Logistical differences in funding sources, payment schedules, health benefits, and tax filing were all cited as common issues. These difficulties were often made even more complicated for international students. One interviewed student summarized their sentiments as:

\begin{quote}
    ``I would like more guidance about money matters. [...] I'm completely unaware of these things, [...] like, how are we getting paid? What kind of contract are we on? Is it an RA, GA, TA, whatever that is? [...] I would love some support." - Dawn
\end{quote}

Dawn is expressing a general confusion toward the financial logistics of being in her program, and is showing that there appears to be a lack of department communication on these matters. The reference to a contract is also referencing confusion about the availability of TA funding after the first year, which was not made immediately clear to students.

\subsection{Valuable Program Aspects: Astrophysics}
\subsubsection{The Program Features a Welcoming Atmosphere and Strong Social Connections Between Students}

The Astrophysics program in this department was described as being welcoming and supportive to their students by the vast majority of collected student data. There is evidence of a significant network of social connections between the students at all levels of the program, which has made for a strong sense of community. One student in the focus group describes their first year experience:

\begin{quote}
    ``I definitely feel like it's been very easy to find a community here. It's been very welcoming and easy to feel involved. Everybody I've interacted with cares a lot, and you can tell.'' - Caroline
\end{quote}

The environment that Caroline is describing is a program culture of students being actively involved and bought-in to the well-being of fellow graduate students in the program. This extends to specifically to students who are members of the LGBTQ+ community, which came up in the focus group several times. 

Another first-year student describes a similar feeling of community among the students in their research group: 

\begin{quote}
    I feel like we have a big group, but it also [has] big family vibes. Everyone's really open to helping each other out. Whether that's like, `when you did this class last year, you know what was going on there?' There's a lot of open communication there, which I appreciate.'' - Sam
\end{quote}

Sam here is describing an example of the extensive amount of advice and support between the senior members of the Astrophysics program and first-year students, most often around academic and research topics. 

The positive views on the culture in the program and overall department can be seen in the qualitative survey responses from students beyond their first year in the program as well, with one student saying:

\begin{quote}
    ``The people here (both faculty and peers) have been incredibly kind and helpful, which has made the environment feel supportive and welcoming.” - Josh
\end{quote}

Another student echoed a similar sentiment:

\begin{quote}
    ``I feel supported and that I am in an extremely healthy and non-competitive program.'' - Jess
\end{quote}

Jess here seems to think she is not in direct competition with her peers, likely for things like research funding or grades in courses. Another student survey response shared this sentiment:

\begin{quote}
    ``We have a very supportive group of people (it is not cut-throat among each other, we help each other). Very happy with the culture and program here!'' - Cameron
\end{quote}

Cameron seems to be implying that an air of competition between students would be detrimental to the culture within the Astrophysics program, and is happy to see this is not the case. 

These student sentiments provide evidence for the overall positive and supportive culture within the Astrophysics program. However, there was one particular student experience that stood out as different from this overall trend. A qualitative survey response described a more negative perception of the program, especially around the culture in the department. This student, referred to with the pseudonym Danielle, described a male-dominated program environment for faculty and students alike, with a need for increasing diversity in hired professors:

\begin{quote}
    ``I have felt confused about the lack of diversity of any kind [among] our faculty, making it hard for me to find mentors and champions in them, and it feels like the department as a whole values their reputation, the opinions of established professors, and money more than the well-being of their students.'' - Danielle
\end{quote}

In this quote, Danielle is expressing her overall disappointment with the culture in the department, especially around faculty reputations superseding the experiences of their students. Conversely, Danielle also described the positive effects that students have had on the program over time:

\begin{quote}
    ``The students really care about the program and have worked to change a lot of things that have benefited future students (i.e. our curriculum, being part of the stipend committee, women in science, being on admissions committee and making sure we cast a wide net and prioritized different kinds of diversity in future student cohorts)'' - Danielle
\end{quote}

Danielle's description of the student body shows that there are many students that have a sense of stewardship over their program and have a desire to see it grow and improve for future cohorts. While the majority of student data that we have collected seem to have positive views on the culture of the program, Danielle's experience shows that this is not universal, and that there is always a possibility of students having negative experiences that go unheard.

\subsubsection{Program Support Structures Toward Academic and Professional Development for Students}

The Astrophysics program seems to have a focus on supporting their students in the transition to graduate school, as well as toward their eventual careers. One program support structure toward this goal is the assigning of two senior peer mentors to incoming first-year students. While these peer mentors do not have any formal requirements from the program, the interviewed first-year students described interacting with their mentors on a regular basis, along with getting frequent advice toward research and required coursework.

The first-year students are also required to take weekly graduate seminar courses in their first two semesters. These seminar courses are designed to develop student skills like reading and writing academic papers and grants in their field, as well as navigating the transition from undergraduate to graduate workloads and expectations. 

These seminars also had sessions dedicated to helping students navigate aspects of being a graduate student that are not often explicitly taught~\cite{hopkins_demystifying_2024}, such as navigating potential difficulties in the relationship with one's research advisor. Other examples of these seminar topics are shown here by Caroline:

\begin{quote}
    ``For professional development in our Grad Sem, it's been probably a pretty even mix. There were a lot of things like impostor syndrome, how to talk to people, how to network, elevator pitches, but then there were a lot of more academic things like `how do you write a paper?' [or] `how do you write a grant proposal?' '' - Caroline
\end{quote}

Here Caroline is discussing a mix between navigating common difficulties for students early in graduate school, having students practice research skills necessary for participating in academia, and professional skills that could prove useful after graduation.

When discussing professional development in their first year in the program, all three students felt that this mostly occurred during their graduate seminar courses, and were generally satisfied with their experience overall. Other opportunities like conferences and meeting researchers at other universities were also discussed, but were specific to the research group or advisor that a student had joined. Some professional development workshops were also available to students as communicated by email by their program director, though the interviewed students felt that they lacked the time to participate in these during their first year. 

\subsubsection{Many Courses Were Described as Well-Designed and Relevant to Research}

All first-year students in the program are required to take the same two graduate courses on the fundamentals of astrophysics, which are offered sequentially in the fall and spring semesters. These fundamental courses were described as positive experiences by the first-year students, citing course content and connections made with relevant research. When discussing these required courses, Caroline said:

\begin{quote}
    ``Pedagogy-wise, I would say I'm happy with how they have fundies formatted, I think that they put a lot of work into that, and it's evident.” - Caroline
\end{quote}

As evidenced by Caroline's thoughts, these fundamental courses have been developed and improved over the last several years, mostly based on student feedback.

Astrophysics students are also required to take several elective courses in their chosen research area. Many of these courses were also described as well-designed and relevant to current research:

\begin{quote}
    ``I think the electives really [are] tailored to research. [I'm taking two electives] and both of them at the beginning were like, `What are your research [areas]? And what are you interested in learning more about? What would help you with your research?' And so we made a list of topics, and they tailored the class [to them].'' - Sam
\end{quote}

Sam here is describing positive experiences with courses centered around interstellar medium and extra-galactic astrophysics, which were designed around the research interests of the students. Tracy had similar experiences in elective classes, though Caroline felt that the courses on relativity and gravity did not share these features.

\subsubsection{Astrophysics Faculty and Program Leadership are Supportive and Respectful of Student Mental Health}

Mental health was a frequent topic of conversation during the focus group interview, most often being brought up in the context of maintaining a healthy work life balance between coursework, teaching duties, research, and personal lives. The three interviewed first-year students all reported getting support in these efforts from many of the program faculty and leadership. When discussing coursework, Caroline said:

\begin{quote}
``I feel like all the professors I've had have been super understanding. Like I don't think I’ve ever asked for an extension and not gotten [one]. They're very much supportive of like `if you need something, you know we will help you the best that we can.' '' - Caroline
\end{quote}

Caroline's is using coursework extensions as an example of faculty being aware of the difficulties associated with the first year in a graduate program, and in turn attempting to ease some of that pressure on their students. Sam describes a similar experience:

\begin{quote}
    ``I feel like it's really easy to be honest with with [my advisor], as to how I'm doing mental health-wise, how I'm doing in classes, where I'm at in terms of projects [...], or not doing as much one week or another.'' - Sam
\end{quote}

Based on this quote, Sam's relationship with their advisor seems to be healthy and supportive despite the large number of students in their research group. 

Program leadership was consistently praised by the interviewed students as being open and available to discuss mental health issues that they faced in their first year in graduate school. All three students described being able to sit with the program director in one-on-one meetings to talk about various issues they were having, with the program director being consistently understanding and supportive while trying to find solutions to their issues. 

Several qualitative survey responses showed that this perception of mental health support among faculty exists beyond the first year in the program. When describing program strengths, one student past their first year said:

\begin{quote}
    ``I truly feel as though I could talk to any faculty member in my program about personal, academic, or research struggles.” - Jess
\end{quote}

Jess is indicating that she feels supported in several ways by a variety of faculty within the department. While this acceptance of mental health for students is not likely to be universal across all faculty in the program or department, the data collected in this project indicate that it is a common trait in many of the faculty that students interact with on a regular basis. 

\subsection{Needed Program Changes or Improvements: Astrophysics}

\subsubsection{Sense of Disconnection Within the Program After the First Year}

While there does seem to be strong social community between the students in the Astrophysics program, this is also negatively affected by a sense of disconnection that builds as students progress through the program. The first year of the program has students in common courses and seminars, resulting in a large amount of time spent with other members of one's cohort. Once these common classes are finished, the students in the program tend to spend larger portions of their time within their own research labs and specialty areas, as shown by this student's thoughts on ways the program could improve:

\begin{quote}
    ``I think they could improve how often there are program-wide events, we tend to separate from other research groups after our first year.'' - Jess
\end{quote}

Here Jess is presenting the idea of the department counteracting the natural social disconnection in the Astrophysics program by putting on official social events on a regular basis. The need for such events is also expressed by a student survey response:

\begin{quote}
    ``No center of mass. Faculty and classes are all over campus. Hallway interactions are non-existent.'' - Louis
\end{quote} 

Louis is referring to the physical separation of not only the different research groups in the Astrophysics program, but of the physics department as a whole, and feels that this is having a negative effect on the informal interactions between students and faculty. Another student survey response expressed a similar view:

\begin{quote}
    ``Our program is split in three areas within astrophysics. [...] While this helps organize our large program, it also fragments it. This prevents us from leveraging opportunities across these fields.'' - Cameron
\end{quote}

Cameron is referring to potentially lost opportunities for collaboration between different research groups within the same department, and is illustrating that the effects of the physical separation of the program may go beyond just the social experiences for the students. This sentiment was shared in another student survey response:

\begin{quote}
    ``Our department is spread out, both in the variety of research fields and physically on campus, and there's not a lot of unity after you finish taking classes, so it's easy to feel isolated.” - Danielle
\end{quote}

Danielle's mention of a sense of isolation is a common experience for students in graduate STEM programs \cite{ali2006dealing,ali2007dealing}, and seems to become more pronounced as the Astrophysics students get further into the program and spend more time in their research groups than with members of the program as a whole.  

\subsubsection{Research Experiences are Varied for Students in the Astrophysics Program}

When applying for the Astrophysics program, students indicate three advisors that they would have interest in working with. Program leadership then assigns one of these three professors as an academic advisor, with the final decision largely based around available funding and advising bandwidth. While this academic advisor is not necessarily who will become the research advisor for a particular student, this appears to be most often the case based on the student data collected in this project. The interviewed first-year students that that the overall process has room for improvement and transparency to prospective students. When discussing the matter, one student said: 

\begin{quote}
``I'm lucky and got paired with a phenomenal advisor [...] who I am sticking with. Sometimes it works out, other times not.'' - Sam
\end{quote}

Sam is recognizing that there are a mix of research advisor relationships for the students in the Astrophysics program. Sam is also referencing how central the role of this relationship is to the overall experience for students throughout their time in pursuing their degree. Caroline and Tracy did not consider themselves as lucky with their paired advisors, with Tracy describing a research group that had an impersonal and rigid atmosphere, with informal conversations with the advisor being rare or even nonexistent. Caroline also expressed thoughts regarding research group assignments:

\begin{quote}
    ``I also think the way research advisors are assigned could benefit from change, because right now it feels very suffocating and daunting if you aren't assigned someone you want to work with.” - Caroline
\end{quote}

This quote illustrates the larger trend of frustrations felt by some Astrophysics students toward the process of attempting to switch to a different advisor or research group. Students were generally able to discuss the prospect of joining another research laboratory with the program director, who was understanding of their desires. However, these students were often told that a lack of available funding would be the main barrier for them being able to change research groups, rather than any explicit department policies.  

\subsubsection{Some Courses Viewed as Not Engaging or Excessively Difficult}

Astrophysics courses were not universally praised by students in the program, with some being described as poorly designed or not very engaging for those taking them. One survey response summarizes these feelings: 

\begin{quote}
    ``The lectures often feel less engaging than I expected. That might just be part of the graduate school experience, but it has made it harder for me to stay fully motivated in my coursework.” - Josh
\end{quote}

While it is not clear if Josh is referring to the required fundamentals courses or electives, it can be inferred that at least some courses in the program are largely taught with a traditional passive lecture model that he does not find very engaging. During the focus group interview, Caroline gave a more specific example with a similar sentiment:

\begin{quote}
    ``The way that [the professor] teaches is very kind of removed, because [there will be] slides that have pictures of a bunch of equations, and then [the professor will] explain one by one, all of the different pictures of equations on [the] slide, and it's just so hard [to] internalize anything.'' - Caroline
\end{quote}

Here Caroline is describing a typical lecture given in an elective course in one of the subfields of astrophysics in the department, where derivations of equations are projected on the board and pointed out one by one, rather than writing them out for students to do so at the same pace. Caroline also expressed an opinion that this particular course had an unrealistic workload of assigned homework, both in terms of quantity and difficulty of problems. Caroline's negative opinions applied to direct experiences in the courses in this subfield, which contrasted with the more positive experiences of the other two interviewed students in their own subfield specific courses. These different experiences show that the courses taken by students in the program are inconsistent in quality and teaching methods.  

\subsubsection{Several Logistical Barriers and Financial Burdens that the Department Could Mitigate}

Several of the program aspects that the students felt needed improvement were separate from course requirements, teaching duties, or research projects typical of the first year in graduate school. The two interviewed students in the Ph.D. program expressed some frustration with their financial situations, starting with the process of moving to the area of the university:

\begin{quote}
    ``I know that other grad schools provide support with moving costs, and that's something that [this school] does not. [...] I had savings from working in undergrad, so it was manageable for me, but it really just ate up all [those] savings.'' - Sam
\end{quote}

Sam moved a considerable distance across the United States to attend this program, and felt that there was little financial support from the department to make this happen. To offset this cost, Sam began taking up part-time jobs like note-taking and delivering food to maintain financial independence while in the program, while simultaneously taking up the small amount of free time that was available. Sam and Caroline both stated that their current stipends as teaching assistants were only able to meet their cost of living, with no money left over to be put into savings. Tracy on the other hand was actively losing money during the first year due to M.S. students paying full tuition, and was only able to make ends meet to to family financial support.

The interviewed students also expressed some confusion around the logistics of financial matters, such as taxes or health benefits:

\begin{quote}
    ``I think it would be helpful if [...] we had more guidance about how to deal with health insurance [or]  taxes. [...] We’re offered health insurance as part of the Ph.D. package, but I have no idea how to set that up.'' - Caroline
\end{quote}

Caroline here is expressing an opinion that current guidance from the department on these matters is insufficient. Students in the program are often left to asking advice from those in the program longer than them. While the advice from more senior students is an unofficial source of support for those in their first year, official guidance from the department on these  logistical matters would be welcomed by the students in the program.

\section{Discussion}
In the above section, the student experiences from the Physics and Astrophysics programs are summarized as emerging themes. These themes fell into the two major categories of Valuable Program Aspects and Needed Program Improvements or Changes. In this section, we will discuss the similarities and differences between the Physics and Astrophysics results, as well as the implications for changes that new Physics Ph.D. program can implement in the future. Implications for the Astrophysics program will be discussed in future work with program leadership, as well as future publications.

\subsection{Limitations}

The quantitative data collected via distribution of the ASES survey tool in this project was limited by the number of 17 complete responses. Thus the sample that was collected was insufficient to conduct validity and reliability measurements of the test scores. 

The nature of focus group interviews presents some potential limitations in the responses of the participants. While efforts were made to create a comfortable environment for the interview, and participants were told the steps taken to ensure the confidentiality of their responses, a group setting can cause some participants to not be as open to talking about sensitive topics like their own mental health or negative experiences they may have gone through. The interviewed students may have also been less willing to voice dissenting opinions in a group setting. 

The focus group that was conducted with first-year students in the Astrophysics program was potentially subjected to self-selection bias. Limiting the focus group interviews to only first-year students was intentional, as this would allow more direct comparisons to the Physics Ph.D. program that was in its first year of existence. This group was made up of two Ph.D. students and one M.S. student from the Astrophysics program. While there were only three first-year M.S. students at the time of data collection, the first-year Ph.D. cohort was made up of ten students. This limited the number of perspectives in our in-depth focus group interview data from this cohort of the Astrophysics program. Two additional first-year students in this category completed surveys and were included in the overall analysis, bringing the total to four out of ten possible student perspectives being recorded to some degree.

The final major limitation in our data set centers around the survey responses from a senior Astrophysics student. This student gave an extensive and detailed qualitative survey response, notable for several negative experiences with the overall culture in their program. These negative experiences were significantly different when compared with other collected survey responses and focus group interviews. We followed up for a one-on-one interview with this student, but this offer was declined. These negative experiences may not be limited to this particular student, and could be reflective of the self-selection bias that was discussed earlier in this section. 

\subsection{Comparisons Between the Physics and Astrophysics Programs}
\subsubsection{Program Culture and Climate}
Both departments were perceived as having positive and supportive overall cultures. Several students in the Astrophysics programs expressed appreciation for a lack of competition between themselves and their peers, a sentiment that was echoed by the Physics Ph.D. students, specifically Chris when he was describing his surprise at what he saw as a healthy physics department. The students seem to expect their department to have an air of competition and individualism, which is a sentiment that has been shown to be the case in prior research on culture in physics and astrophysics programs~\cite{ong_counterspaces_2018}. The sense of community and belonging for members of the LGBTQ+ community in the Astrophysics program is a significant result in that it differs from documented experiences of these students in other physics and astronomy graduate programs~\cite{barthelemy2022lgbt+}. While this result was not found in the student data from the Physics program, this does not necessarily imply that this program is unwelcoming to those students with these identities. The overall culture in both programs appears to generally be welcoming and supportive, they would both benefit from emphasizing and reinforcing this feature of the department, as this could be beneficial to the experiences and program persistence for any current or future students with underrepresented identities in STEM~\cite{ali2006dealing,golde2005role}.

\subsubsection{Student Social Experiences}
The social experience between the peers of the first Physics Ph.D. cohort showed some significant differences when compared with those of the first-year students in the three other graduate programs in this department. The Physics Ph.D. cohort reported a relative lack of connections with their peers when compared to the strong social network between the first-year students in the Astrophysics program. A significant contributor to this effect was likely the lack of common courses shared by this first cohort, as six out of the seven students transferred graduate course credits from prior institutions. The Physics M.S. students seemed to have more social connections than their Ph.D. counterparts, which may have resulted from them sharing these common courses. The ability to transfer graduate credits is a program feature that this first Physics Ph.D. cohort highly valued, and should be kept for future cohorts. However, more social events for these graduate students will be needed to supplement the building of community for future students, as avoiding social isolation should be a priority for the program~\cite{sachmpazidi_role_2025}.

Another key finding is the sense of disconnection felt by the Astrophysics students as they progressed in their program. This seemed to be mainly caused by the increased time spent in the physically separated research areas in the program, as well as a lack of collaboration between these lab groups. 

\subsubsection{Student Mental Health}
Both programs seem to have informal support systems in place in support of the mental health of their students, though formal policies and resources appear to be lacking within the department as a whole. Research mentors were the main source mentioned by the Physics Ph.D. students, while the Astrophysics students indicated program leadership and their teaching professors as their own sources of support. Students from both programs expressed a desire for more formalized resources around mental health support, though this was a more consistent theme for the Physics Ph.D. students, who discussed the currently existing support system in the counseling available through the student health center, though this has the drawback of costing money out of pocket for students who want to make use of it as a mental health resource.

\subsubsection{Coursework}
Both the Physics and Astrophysics programs have a mix of common core courses and more flexible elective courses that are required of their students. The Physics Ph.D. students expressed some frustrations with the number of required core courses, the amount of time needed to dedicate to them, and their relevance to their research. These feelings were not entirely shared with the Physics M.S. students, who generally enjoyed these required courses. The Astrophysics students only had two required core courses, compared to the four for Physics students, and these courses were found to be generally well-structured and relevant to the field of Astrophysics. Some of the elective courses for the Astrophysics students were viewed as relevant to chosen fields of research, though this was not consistent across the board. All courses for all programs were generally taught in traditional lecture format, with few instances of active learning or student-centered pedagogy being implemented.  

\subsubsection{Research and Mentoring}
Both the Physics and Astrophysics Ph.D. programs have prospective students pick out potential advisors and research areas of interest as a part of their application process. This policy may be a significant contributing factor to the high rate of students having research advisors in their first year of their programs. Finding a research group can be stressful and inequitable for Ph.D. students~\cite{verostek2023inequities,verostek2024physics,verostek2024modeling}, and departmental supports may help alleviate those issues, such as the Astrophysics Ph.D. program assigning academic advisors with available lab space and funding to incoming students. The interviewed students did express some frustrations with the transparency of this process. Accepted students were not given a timeline for these decisions, nor were they told which faculty were deciding final placements. There were then subsequently some barriers to exploring or switching lab groups for students who were not happy with their research experiences. These concerns were not generally shared by those in the Physics Ph.D. program. This does not necessarily reflect the experiences of all Astrophysics Ph.D. students in this program, as selection bias in the focus group interviews may have affected the results.

The Physics Ph.D. students described interdisciplinary and collaborative research groups and opportunities, which was a feature that was not as present for the Astrophysics students. This seems to be a symptom of the siloed nature of the three research subdisciplines within the Astrophysics program. 

All of the graduate students in our data set have received some form of mentoring from their research advisor. However, the students in the Astrophysics program also had the benefit of senior peer mentors assigned to them through the department in their first year. While this program had no formal requirements of the mentors, it does provide an avenue for first-year students to get useful advice on many issues related to the program, including coursework, research, and navigating logistical matters. The Physics Ph.D. students did not have such a structure in place to give them an additional mentoring source, though the early nature of the program rendered this largely unfeasible to implement for the first cohort. 

\subsubsection{Professional Development}
Both programs are lacking in professional development opportunities and guidance, especially for career paths outside of academia. This is consistent with prior findings in graduate physics education research~\cite{sachmpazidi2021departmental,sachmpazidi2025beneath}. Both the Physics and Astrophysics programs placed most of their professional development efforts for their first-year students in their respective graduate seminar courses. The Astrophysics graduate seminar seems to be more successful in these endeavors when compared to its Physics counterpart, with time being spent on skills like reading and writing grant proposals and navigating the relationship with one's research advisor. 

Within our data set, the students in the Physics program had consistently more positive and supportive relationships than their counterparts in the Astrophysics program. One possible explanation for this is the current small size of the Physics Ph.D. program relative to the Astrophysics Ph.D. program, meaning that advisors have more time and bandwidth to give individual attention and mentoring to their research students when compared to the larger Astrophysics research groups. This may change over time as the Physics Ph.D. program grows in size with new cohorts of students in subsequent years.

\subsubsection{Logistics and Department Communications}
Students in all three conducted focus groups expressed some level of frustration with the department's communication of logistical issues. Communication before arriving on campus was a common theme in this category, with physics students being confused about what GTA opportunities were available and their duration, while the Astrophysics students felt that explanation of the process of exploring and changing research groups was miscommunicated. Students in both programs were generally confused on financial matters like setting up health benefits or filing taxes when receiving funding from multiple sources.

\section{Presenting Analysis To Department Leadership}
\subsection{Initial Meetings}
Once the analysis of both the quantitative and qualitative data was completed, the researchers set up initial meetings with faculty in leadership positions in each program. The first of these meetings was with the director of the Physics Ph.D. program, with the second being with the assistant director of the Astrophysics Ph.D. program. Both of these meetings involved presenting a slide show with relevant findings to each respective program, along with an ensuing conversation to address questions.

The faculty leadership from both programs expressed an eagerness to hear directly from students regarding their views and experiences. For the Physics Ph.D. program, the director recognized the importance of this kind of data collection early in the program's existence. It was also found that some of the issues and concerns of students were already being addressed. For example, many of the logistical confusion felt by the first cohort may be alleviated by the online digital graduate student handbook that had been created and distributed. The director had also been making efforts to build a social community with regular extracurricular events for students, along with a shared Slack channel for updates and communication.

The meeting with faculty leadership for the Astrophysics program followed a similar trend overall. The assistant Ph.D. program director expressed an appreciation for the communication of student experiences, though expressed an opinion that the student feelings were not particularly surprising based on direct interactions with the students themselves. 

\subsection{Action Items}
Both program leaders accepted an offer from the researchers to compose a list of action items for each program to potentially implement toward addressing student concerns and issues. The researchers composed two distinct lists for the respective programs, though they shared overall similar content and structure. The full lists given to program leadership can be viewed in the Supplemental Material. Brief summaries of each category of action items are given below:

\begin{enumerate}
\item \textit{Mental Health and Student Well-Being}: Strengthen mental health resources and community support. The department could establish new structures like regular advisor-student meetings dedicated to well-being check-ins, or having a dedicated staff mental health liaison. The department could also clarify and communicate current resources offered through the university, such as counseling through the student health center. 
\item \textit{Academic Support and Teaching Readiness}: Improve academic support structures and teaching readiness. This would involve the creation of formalized support plans for any students who struggle in their core coursework, such as tutoring or study groups led by senior graduate students in the program. For those students who are working as GTAs, a training program is needed that focuses on teaching skills, grading, and communication skills.
\item \textit{Financial and Administrative Guidance}:  Improve clarity around funding, benefits, and logistics. First-year students would benefit from a training session during their orientation on the financial aspects of being a graduate student, including setting up health benefits and filing taxes based on income sources. The creation of flowchart around these financial obstacles would be useful for students later on in the program as well.
\item \textit{Research and Mentoring}: Capitalize on a strength of the program by improving mentoring and support for students even further. Possible actions include starting and formalizing a multi-mentoring model for students. Example roles in this model consist of the existing research mentor, an academic advisor offering additional support toward academic progress and program milestones, and a senior peer mentor who can assist in navigating coursework and department culture. A framework for high-quality mentoring should be created then distributed to those in mentorship roles.
\item \textit{Community Building}: Strengthen peer connections and a sense of community and belonging. The department should continue the monthly social events organized by the program director, as well as help facilitate student-organized events like regular movie nights. Finding communal desk space for first-year students should also be made a priority.
\item \textit{Professional Development}: Broaden preparation for careers both inside and outside academia. Increased efforts should be made to have alumni from the department return as guest speakers for colloquia presentations, especially those who have made careers outside of academic institutions. A formal plan for each student based on the Individualized Development Plan (IDP) from AAAS can also be implemented \cite{myIDP_AAAS_2025}.
\item \textit{Curriculum and Faculty Teaching}: Align coursework with students’ research interests. Program leadership should consider creating tracks aligned with chosen research areas for the Physics Ph.D. degree. This would allow students to fill more of their degree requirements with more courses that they consider relevant to their research field, while still maintaining some general physics courses in common for all students in the program. Also consider implementing more active learning elements (e.g., class discussions) in courses traditionally taught with passive lectures.
\item \textit{Program Communication and Transparency}: Ensure clear, consistent communication of program expectations. The department can help in this category by keeping their graduate handbook up to date and easily accessible for all students in the program. The program can also communicate program expectations and milestones well before accepted students arrive on campus. 
\item \textit{Transition to Graduate School}: Support graduate students in navigating a new program and social environment, especially for those who are coming directly from their undergraduate experience. The Astrophysics program offers several examples that could be implemented in the Physics program, including establishing a peer mentor program between senior and first year students, or reviewing and adapting successful practices from the Astrophysics graduate seminar courses (This item was only given to the Physics Ph.D. program director). 
\item \textit{Graduate Seminar Structure}: Enhance the impact of graduate seminar courses. The graduate seminar courses should be refocused around practical research and professional skills for the students. Example topics include Resume/CV creation, job searching and mock interviews, reading and writing grant proposals, making posters and presentations for relevant conferences, and learning about the notable researchers and lab groups in one's field (This item was only given to the Physics Ph.D. program director).
\end{enumerate}

\subsection{Ongoing Departmental Discussion}

Both program leaders were sent the action items for their respective programs and given several weeks to read and process the proposed changes. The researchers then reached out again to elicit feedback from the program leaders on the proposed action items. The requested feedback centered around the program leaders' views on feasibility and time scale for implementation, as well as perceived institutional or cultural barriers.

A common barrier toward program change in higher education that was directly experienced in this research project was that of available time and coordination between all parties. Both program directors had many administrative, teaching, and research duties to attend to during the academic year, so time to spend on responding to researcher emails and meetings was limited. This also contributed to delays in making changes in either program.

While neither program director has been able to make time for further face-to-face meetings or discussions as of the time of writing, both were able to give their feedback and perspectives asynchronously. According to the Astrophysics Ph.D assistant director, several of the action items were already being addressed, including having multiple mentors for students, a dedicated graduate handbook, and spending graduate seminar time on the logistics like health insurance or taxes. The assistant director also acknowledged that some of the logistical issues may need to be emphasized more or with a different approach, especially if there are still some students who are still confused or not aware of the resources at their disposal. A common barrier mentioned in the feedback was a lack of financial resources for needed social events, and that coordination between the two Ph.D. programs may be beneficial on this front. While mental health was generally supported by program faculty in the eyes of the interviewed students, the assistant director also felt that having a dedicated staff liaison for students on this front was not feasible with current staff resources. Increased use of active learning techniques in graduate courses was highlighted as a good idea, and was reported to be happening in some courses, but the assistant director also felt that a wider cultural shift on the part of program faculty was necessary.

The Physics Ph.D. program thus far has not been able to give the same level of detailed feedback to the researchers based on limited time during the semester, but mentioned that they wanted to give their full attention to the proposed program changes, as well as discussing them with the faculty on the Physics Ph.D. program committee and the overall department chair to elicit their ideas and feedback as well in early Fall 2026. Working toward changes and improvements to student experiences and outcomes for both of these graduate programs will be an ongoing and iterative process that will require sustained effort from all parties involved. 

\section{Implications for Practice and Research}

The primary goals of this study were to document the experiences of the graduate students at a single physics department, then work with program leadership to begin a cycle of reflective practice toward improving those student experiences over time. This inherently means that the results and analysis presented are specific to this one context, and are not able to describe graduate student experiences in other graduate programs. However, there is still a significant amount of insight that can be gained for researchers and practitioners in the field of graduate education. 

In this paper we have described a model for laying the groundwork for collecting student data toward an iterative cycle of program improvement in a newly established physics graduate program. Graduate programs would benefit from implementing the research practices laid out in this paper. Systemic change in graduate education is a slow and difficult process, and meaningful and long-lasting change requires deliberate and sustained action on the part of program leadership in conjunction with data collected and analyzed by researchers or practitioners~\cite{eckel_taking_2011,kezar_how_2018,sachmpazidi2022changing}. This paper describes an approach for starting a collaborative effort for systematic data collection in local departments. 
A detailed description of these research methods will be made available in digital format to interested researcher and practitioners upon request to the authors.

\section{Conclusions}

STEM graduate programs have a longstanding history of creating unsupportive, individualistic, and exclusionary environments for their students~\cite{oakes_chapter_1990,ali2006dealing,golde2005role,ong_counterspaces_2018}, which can lead to negative student experiences that result in leaving a program before graduation~\cite{lovitts2001leaving,Sowell2015}. Change efforts in this field have had varying results~\cite{apsbridge,igenetwork,goldberg2024inclusive}, with many of these initiatives focusing only on program structures and not the culture within the program itself~\cite{ong_counterspaces_2018,cochran_understanding_2025}. 

This paper documents the efforts of the researchers to collaborate with the leadership of a newly established Physics Ph.D. program in order to foster a culture of regular data collection on student experiences. Quantitative and qualitative student data were collected from both the new physics program and the preexisting astrophysics program in the same department. These data were used to paint a picture of the experiences of the first cohort of the Physics Ph.D. program and their counterparts in the Astrophysics program. Students from both programs highlighted aspects of their programs that they found valuable, as well as what they saw as needed program changes or improvements, and the results from both programs were compared with each other. These results and analysis were presented to the director of the Physics Ph.D. program, and a list of actionable changes and points of emphasis for the program was compiled in a collaborative process.

The work presented in this paper represents the first steps of long-term change efforts by the researchers within the department under study. The most immediate steps to be taken are to follow up with the Physics Ph.D. program director to discuss the decision-making process around implementing action items in both short-term and long-term time frames. Similar meetings to discuss results and analyses with the faculty leadership for the Astrophysics program are planned for the future. Further iterations of this data collection process and program reflection are planned as well as part of the cyclic and sustained nature of these change efforts.

Researchers and faculty leadership interested in institutional change efforts for their STEM graduate programs could find significant benefit in following the model of data collection and analysis depicted in this paper. This could potentially allow for the discovery of hidden graduate student experiences in a wide range of academic contexts, as well as a way to begin the process of improving those experiences, and in turn create meaningful and long-lasting change for future STEM graduate students.

%

\clearpage
\appendix
\section{Comparing ASES Results Between Degree Programs}

\begin{table}
\centering
\begin{ruledtabular}
\caption{\label{tab:ASES_comparison_MS_PhD}%
ASES factor comparisons between responses from MS programs and Ph.D. programs in the department under study. Hedge's $g$ values were used due to small sample sizes (N = 5 for M.S., N = 12 for Ph.D.)}
\begin{tabular}{lcc|cc|c}
Factor 
& \multicolumn{2}{c|}{M.S.}  
& \multicolumn{2}{c|}{Ph.D.} \\
\hline
& Mean & SD  
& Mean & SD & $g$ \\
\hline
MRE & 4.23 & 0.23 & 4.20 & 0.49 & 0.06 \\
PD  & 2.69 & 0.76 & 2.83 & 0.78 & -0.16 \\
SAI & 3.12 & 0.32 & 3.35 & 0.67 & -0.37 \\
FS  & 2.20 & 0.61 & 3.88 & 0.72 & -2.30 \\
\end{tabular}
\end{ruledtabular}
\end{table}

\begin{table}
\centering
\begin{ruledtabular}
\caption{\label{tab:ASES_comparison_Phys_Astro}%
ASES factor comparisons between responses from Physics programs and Astropyhysics graduate programs in the department under study. Hedge's $g$ values were used due to small sample sizes (N = 8 for Physics, N = 9 for Astrophysics)}
\begin{tabular}{lcc|cc|c}
Factor 
& \multicolumn{2}{c|}{Physics}  
& \multicolumn{2}{c|}{Astrophysics} \\
\hline
& Mean & SD  
& Mean & SD & $g$ \\
\hline
MRE & 4.32 & 0.25 & 4.12 & 0.53 & 0.44 \\
PD  & 2.40 & 0.87 & 3.14 & 0.43 & -1.03 \\
SAI & 3.16 & 0.83 & 3.39 & 0.69 & -0.36 \\
FS  & 2.63 & 0.49 & 4.06 & 0.68 & -1.81 \\
\end{tabular}
\end{ruledtabular}
\end{table}

Further analysis of the collected Aspect of Student Experiences Scale (ASES) survey data are presented in this section. 17 out of a possible 63 graduate students in the physics department under study completed the ASES survey, and the mean values for each factor are shown in Table~\ref{tab:ASES_comparison}, along with the same results from two prior distributions of the ASES at other U.S. institutions~\cite{sachmpazidi2021departmental,sachmpazidi2025beneath}. 

These mean factor values were also analyzed by several other variables in the population of graduate students that responded to the survey. The first of these comparisons was by students who were in M.S. programs against those in Ph.D. programs, as shown in Table~\ref{tab:ASES_comparison_MS_PhD}. The low Hedge's g effect sizes suggest that the mean factor values are consistent between the M.S. and Ph.D. students, except for the significant drop in Financial Support (FS) for the M.S. students. This is likely explained by the fact that students in both Physics M.S. and Astrophysics M.S. programs pay full tuition to be in their program, while their counterparts in the Ph.D. programs generally have tuition covered by GTA or RA funding.

The second variable that was analyzed for differences in mean factor value was by students in Physics programs against those in Astrophysics programs, the results of which can be seen in Table~\ref{tab:ASES_comparison_Phys_Astro}. Both Mentoring and Research Experiences (MRE) and Social and Academic Integration (SAI) have small effect sizes between them, indicating similar overall experiences between the two programs. There is once again a large effect size between the programs for Financial Support (FS), but this is once again likely explained by the presence of M.S. students in the Physics sample, while the Astrophysics sample is entirely made up of Ph.D. students. Professional Development (PD) also had a large effect size, indicating that the Physics program may be performing weaker in this factor compare with Astrophysics. While the sample size is too small to make any definitive claims from this analysis, this conclusion is supported by the thematic analysis of the quantitative data collected in this project, which is discussed in the main text of this paper.

\end{document}


\section{Supplemental Material: Department Action Items}
This supplement to our main paper contains the action items given to the department leadership of both the Physics Ph.D. and Astrophysics Ph.D. programs. These action items can be seen in Table~\ref{tab:Action Items Chart}, and are shown verbatim as they were given to the faculty leadership. Several categories were identical between the two programs, as indicated by the first three rows. The rest at least had some differences based on the respective programs, with the last two being specific to the Physics Ph.D. program.

\begin{table*}[]
\begin{adjustbox}{max width=\textwidth, max height=\textheight}
\scriptsize
\begin{tabular}{|l|ll|}
\hline
Category &
  \multicolumn{1}{c|}{Action items: Physics Ph.D.} &
  \multicolumn{1}{c|}{Action Items: Astrophysics Ph.D.} \\ \hline
\begin{tabular}[c]{@{}l@{}}Mental Health \\ and Student \\ Well-being\end{tabular} &
  \multicolumn{2}{l|}{\begin{tabular}[c]{@{}l@{}}\textbullet~Establish departmental mental health support  (e.g., a dedicated staff liaison).\\ \textbullet~Clarify and promote existing university mental health resources; ensure access for graduate students.\\ \textbullet~Implement regular advisor-student well-being check-ins.\end{tabular}} \\ \hline
\begin{tabular}[c]{@{}l@{}}Academic \\ Support and \\ Teaching \\ Readiness\end{tabular} &
  \multicolumn{2}{l|}{\begin{tabular}[c]{@{}l@{}}\textbullet~Create formalized support plans for students struggling in core courses (e.g., tutoring, study groups)\\ \textbullet~Develop and implement a TA training program focused on teaching, grading, and communication.\end{tabular}} \\ \hline
\begin{tabular}[c]{@{}l@{}}Financial and \\ Administrative\\ Guidance\end{tabular} &
  \multicolumn{2}{l|}{\begin{tabular}[c]{@{}l@{}}\textbullet~Include training in orientation on topics like finances, filing taxes, and health benefits\\ \textbullet~Develop a comprehensive flowchart or guide covering GRA vs GTA funding structures, health insurance setup,\\ and hidden fees\end{tabular}} \\ \hline
\begin{tabular}[c]{@{}l@{}}Research and \\ Mentoring\end{tabular} &
  \multicolumn{1}{l|}{\begin{tabular}[c]{@{}l@{}}\textbullet~Take steps towards implementing a multi-mentoring model,\\ including:\\ - Primary Research Advisor\\ - Academic Advisor: faculty who offers additional\\ support toward student academic progress and\\ program milestones, professional development,\\ and student well-being\\ - Peer Mentors: assist students in coursework,\\ research challenges, and department culture\\ \textbullet~Formalize mentorship roles and make them\\ transparent and visible\\ \textbullet~Create a framework for high quality mentoring to give\\ to faculty advisors\end{tabular}} &
  \begin{tabular}[c]{@{}l@{}}\textbullet~Take steps towards implementing a multi-mentoring model,\\ including:\\ - Primary Research Advisor\\ - Academic Advisor: faculty who offers additional\\ support toward student academic progress and\\ program milestones, professional development,\\ and student well-being\\ - Peer Mentors: continue existing mentoring program\\ \textbullet~Formalize mentorship roles and make them\\ transparent and visible\\ \textbullet~Create a framework for high quality mentoring to give\\ to faculty advisors\\ \textbullet~Clarity on the prospects of being able to explore or\\ switch lab groups and/or advisors\end{tabular} \\ \hline
\begin{tabular}[c]{@{}l@{}}Community \\ Building\end{tabular} &
  \multicolumn{1}{l|}{\begin{tabular}[c]{@{}l@{}}\textbullet~Continue monthly Grad Physics social events (led by\\ program director, first Saturday of each month)\\ \textbullet~Launch student-led regular Friday movie nights with a\\ rotating sign-up for movie selection and snacks\\ \textbullet~Maintain or expand common desk space for first-year\\ Ph.D. students\end{tabular}} &
  \begin{tabular}[c]{@{}l@{}}\textbullet~Maintain social connections between Astrophysics\\ students beyond their first year\\ - Casual, recurring events (weekly lunches, movie nights, etc.)\\ - Survey to graduate students asking what kind of\\ events they would like to see implemented, and\\ what factors would make them want to attend\\ \textbullet~Build stronger community between Astrophysics and\\ Physics programs (Schedule  meeting between faculty\\ leadership of both programs to start)\\ Potentially pool resources and ideas for student-led\\ events between both student bodies\end{tabular} \\ \hline
\begin{tabular}[c]{@{}l@{}}Professional \\ Development\end{tabular} &
  \multicolumn{1}{l|}{\begin{tabular}[c]{@{}l@{}}\textbullet~Invite alumni and industry professionals as\\ colloquium speakers\\ \textbullet~Organize career Q\&A sessions with industry speakers\\ \textbullet~Introduce an Individualized Development Plan (IDP)\\ process for all graduate students\end{tabular}} &
  \begin{tabular}[c]{@{}l@{}}\textbullet~Invite alumni and industry professionals as\\ colloquium speakers\\ \textbullet~Organize career Q\&A sessions with industry speakers\\ \textbullet~Introduce an Individualized Development Plan (IDP)\\ process for all graduate students\\ \textbullet~Reciprocal practice talks between graduate student lab\\ groups at other universities\end{tabular} \\ \hline
\begin{tabular}[c]{@{}l@{}}Curriculum \\ and \\ Faculty \\ Teaching\end{tabular} &
  \multicolumn{1}{l|}{\begin{tabular}[c]{@{}l@{}}\textbullet~Evaluate creating specialized Ph.D. tracks\\ \textbullet~Maintain common foundational courses before\\ specialization\end{tabular}} &
  \begin{tabular}[c]{@{}l@{}}\textbullet~Implement more active learning techniques in\\ graduate coursework\\ \textbullet~Begin a community of practice between Astrophysics\\ professors to share effective teaching techniques, \\ learning assessments, research topic incorporation, etc.\end{tabular} \\ \hline
\begin{tabular}[c]{@{}l@{}}Program \\ Communication \\ and Transparency\end{tabular} &
  \multicolumn{1}{l|}{\begin{tabular}[c]{@{}l@{}}\textbullet~Physics Graduate Student Handbook \\ \textbullet~Provide detailed pre-arrival communication outlining\\ program milestones and timelines, TAing expectations\\ (including summer teaching)\\ \textbullet~Keep documentation up to rate and easily accessible\end{tabular}} &
  \begin{tabular}[c]{@{}l@{}}\textbullet~Provide detailed pre-arrival communication outlining\\ program milestones and timelines, TAing expectations\\ (including summer teaching)\\ \textbullet~Clarify advisor pairing process and timeline for\\ prospective students (decision dates, key personnel, etc.)\\ \textbullet~Keep documentation up to date and easily accessible\\ (e.g., online graduate handbook)\end{tabular} \\ \hline
\begin{tabular}[c]{@{}l@{}}Transition \\ to Graduate \\ School\end{tabular} &
  \multicolumn{1}{l|}{\begin{tabular}[c]{@{}l@{}}\textbullet~Establish a graduate peer mentoring program, pairing\\ senior and first-year students.\\ \textbullet~Review and adapt successful practices from the\\ Astrophysics Grad Seminar course for Physics.\\ - E.g. “How to talk to your advisor”, “work-life balance”, etc.\end{tabular}} &
  N/A \\ \hline
\begin{tabular}[c]{@{}l@{}}Graduate \\ Seminar \\ Restructure\end{tabular} &
  \multicolumn{1}{l|}{\begin{tabular}[c]{@{}l@{}}\textbullet~Redesign Graduate Seminar II to focus more on\\ practical and professional skills, including: \\ - Resume/CV creation\\ - Job searching and mock interviews (Academia and Industry)\\ - Reading/writing grant proposals\\ - Making posters and presentations for conferences in your field\\ - Learning about the notable researchers and lab\\ groups in your field\end{tabular}} &
  N/A \\ \hline
\end{tabular}%
\end{adjustbox}
\caption{}
\label{tab:Action Items Chart}
\end{table*}